\documentclass[times,twocolumn,final]{elsarticle}

\usepackage{medima}
\usepackage{framed,multirow}

\usepackage{amssymb}
\usepackage{latexsym}

\usepackage{url}
\usepackage{xcolor}
\usepackage{color}
\usepackage{hyperref} 
\usepackage{booktabs,caption}
\usepackage{makecell}
\usepackage{amsmath,amsfonts}
\usepackage{comment}
\usepackage{graphicx}
\newcommand{\ie}{\textit{i}.\textit{e}.}

\definecolor{newcolor}{rgb}{.8,.349,.1}

\newcommand{\mia}[1]{{\color[rgb]{0,0,0}{#1}}}

\journal{Medical Image Analysis}

\begin{document}

\verso{Shuhan Li \textit{et~al.}}

\begin{frontmatter}

\title{Vessel-Promoted OCT to OCTA Image Translation by Heuristic Contextual Constraints}

\author[1]{Shuhan \snm{Li}}
\cortext[cor1]{Corresponding author: Email: eexmli@ust.hk}
\author[2]{Dong \snm{Zhang}}
\author[2,3]{Xiaomeng \snm{Li}\corref{cor1}}
\author[4]{Chubin \snm{Ou}}
\author[5]{Lin \snm{An}}
\author[6]{Yanwu \snm{Xu}}
\author[7]{Weihua \snm{Yang}}
\author[8]{Yanchun \snm{Zhang}}
\author[2]{Kwang-Ting \snm{Cheng}}

\address[1]{Department of Computer Science and Engineering, The Hong Kong University of Science and Technology, Hong Kong, China.}
\address[2]{Department of Electronic and Computer Engineering, The Hong Kong University of Science and Technology, Hong Kong, China.}
\address[3]{HKUST Shenzhen-Hong Kong Collaborative Innovation Research Institute, Futian, Shenzhen, China.}
\address[4]{Weizhi Meditech (Foshan) Co., Ltd, China.}
\address[5]{Guangdong Weiren Meditech Co., Ltd, China.}
\address[6]{South China University of Technology, and Pazhou Lab, China.}
\address[7]{Shenzhen Eye Institute, Shenzhen Eye Hospital, Jinan University.}
\address[8]{Department of Ophthalmology, Shaanxi
Eye Hospital, Xi’an People’s Hospital (Xi’an Fourth Hospital), Affiliated
People’s Hospital of Northwest University, Xi’an, P.R. China.
}
\received{1 May 2013}
\finalform{10 May 2013}
\accepted{13 May 2013}
\availableonline{15 May 2013}
\communicated{S. Sarkar}

\begin{abstract}
Optical Coherence Tomography Angiography (OCTA) is a crucial tool in the clinical screening of retinal diseases, allowing for accurate 3D imaging of blood vessels through non-invasive scanning. However, the hardware-based approach for acquiring OCTA images presents challenges due to the need for specialized sensors and expensive devices.
In this paper, we introduce a novel method called TransPro, which can translate the readily available 3D Optical Coherence Tomography (OCT) images into 3D OCTA images without requiring any additional hardware modifications.
Our TransPro method is primarily driven by two novel \mia{ideas} that have been overlooked by prior work.
The first \mia{idea} is derived from a critical observation that the OCTA projection map
is generated by averaging pixel values from its corresponding B-scans along the Z-axis. 
Hence, we introduce a hybrid architecture incorporating a 3D adversarial generative network and a novel \textbf{H}euristic \textbf{C}ontextual \textbf{G}uidance \textbf{(HCG)} module, which effectively maintains the consistency of the generated OCTA images \mia{between 3D volumes and projection maps}.
\mia{The second idea is to improve the vessel quality in the translated OCTA projection maps.} As a result, we propose a novel \textbf{V}essel \textbf{P}romoted \textbf{G}uidance \textbf{(VPG)} module to enhance the attention of network on retinal vessels. 
Experimental results on two datasets demonstrate that our TransPro outperforms state-of-the-art approaches, with relative improvements around 11.4\% in MAE, 2.7\% in PSNR, 2\% in SSIM, 40\% in VDE, and 9.1\% in VDC compared to the baseline method. The code is available at:~\href{https://github.com/ustlsh/TransPro}{https://github.com/ustlsh/TransPro}.
\end{abstract}


\end{frontmatter}



\section{Introduction} 
\label{1:intro}
\begin{figure}[t]
\centering
\includegraphics[width=.49\textwidth]{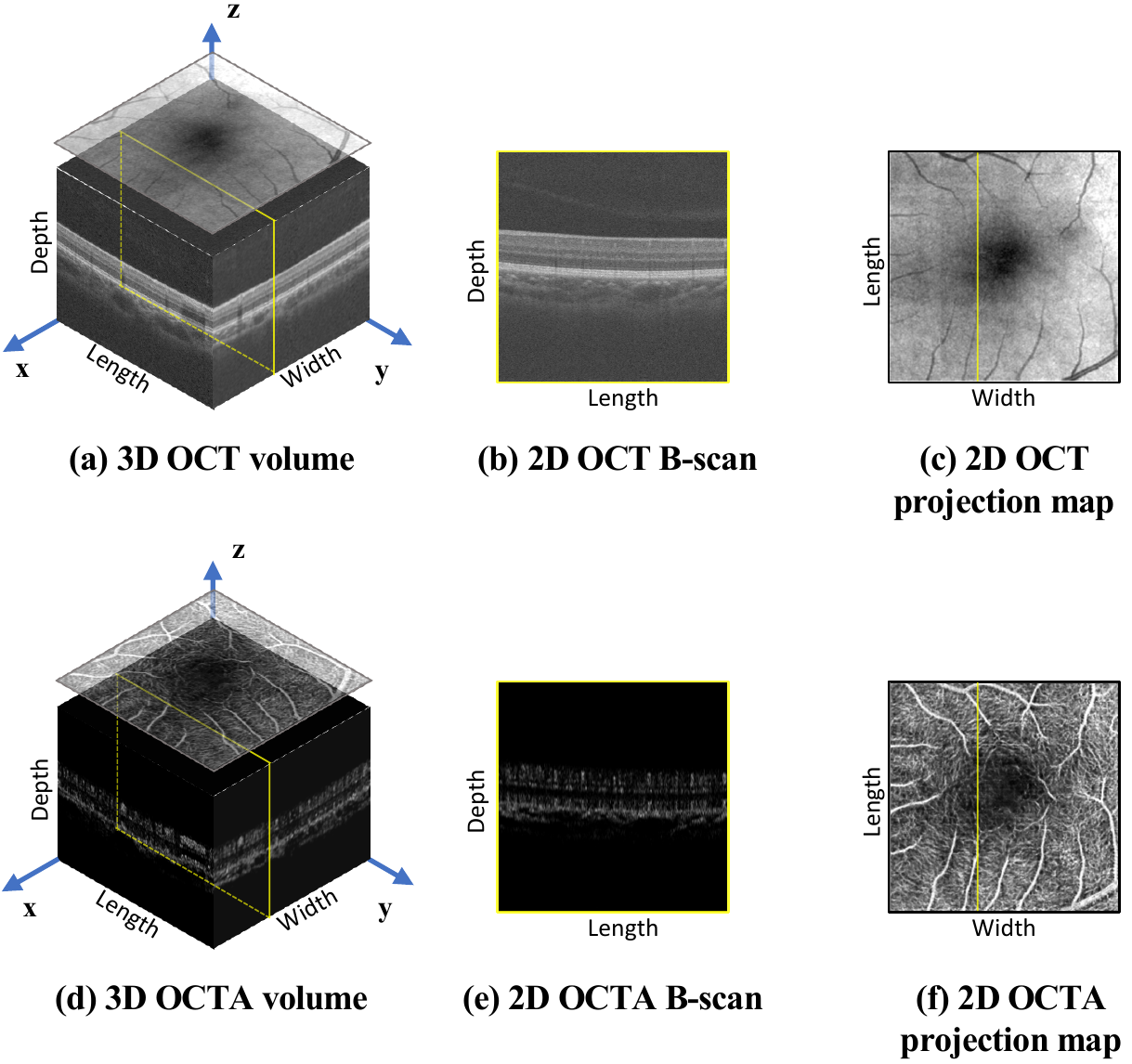}
\caption{Examples of OCT images (in (a)--(c)) and OCTA images (in (d)--(e)). 3D OCT volume (in (a)) or 3D OCTA volume (in (d)) consists of sequential 2D OCT or OCTA B-scan images. \mia{2D B-scan images, which are one slice of the 3D volumes, are shown in (b) and (e).} In addition, by computing the mean of pixel values, the 3D volume can be projected along \mia{Z-axis (the Depth direction)} to obtain 2D projection map. The 2D OCT and OCTA projection maps are displayed in (c) and (f), respectively.
}
\vspace{-4mm}
\label{fig0}
\end{figure}

Optical coherence tomography (OCT) is a non-invasive imaging technique that provides three-dimensional cross-sectional visualization of retinal structures~\cite{huang1991optical}. It plays a crucial role in the diagnosis of various eye diseases, including age-related macular degeneration, diabetic retinopathy, and glaucoma~\cite{fujimoto2016development}.
Building upon the OCT imaging technique, an advanced modality, OCT angiography (OCTA) was developed to display blood flow in retinal microvasculature. OCTA images are obtained by repeatedly scanning the OCT sectional images at the same location of the retina, allowing for the detection of dynamic signals resulting from the movement of erythrocytes within the vascular system~\cite{kashani2017optical}. The availability of OCTA images has been extremely useful in the early detection of some eye disorders, such as choroidal neovascularization, which occurs in wet age-related macular degeneration~\cite{chen2021diagnostic}.
Figure~\ref{fig0} illustrates the 3D volumes of OCT and OCTA, which are depicted in (a) and (d), respectively.
The B-scan image, shown in (b) and (e), represents a single slice within the 3D volume. On the other hand, the projection map, displayed in (c) and (f), is generated by averaging the pixel values along the z-axis of the 3D volume.

Acquiring OCTA images requires additional hardware and software modifications to mitigate the impact of involuntary tissue motions caused by heartbeat, respiration, and eye movement. The high cost of the motion tracking module has resulted in a lower adoption rate of OCTA compared to OCT.
To address this challenge, deep learning methods can serve as alternative approaches for obtaining the cumbersome and costly OCTA modality from the readily available and cost-effective OCT modality.
Several studies have proposed various approaches for OCT to OCTA translation. One common approach is to use a UNet~\cite{ronneberger2015u} as the backbone and employ generative-adversarial learning frameworks for pixel-to-pixel image translation~\cite{isola2017image, zhu2017unpaired}. These existing methods can be categorized into two groups based on the type of input images they utilize.
The first group of methods~\cite{lee2019generating, zhang2021texture, li2020deep} focuses on translating 2D OCT B-scan images into their corresponding OCTA B-scan images. For instance,~\cite{lee2019generating} introduces a deep learning method for OCT to OCTA translation, utilizing an encoder-decoder structure to generate 2D OCTA B-scan images from paired 2D OCT B-scan images. More recent methods incorporate additional information, such as texture features~\cite{zhang2021texture} or adjacent B-scans~\cite{li2020deep}, to enhance the quality of the translated OCTA B-scan images.
The second group of methods~\cite{pan2022multigan} operate on 2D OCT projection maps and generate corresponding 2D OCTA projection maps.

However, the existing OCT to OCTA translation methods suffer from two limitations. Firstly, these methods focus on either B-scan images or projection maps. As illustrated in Figure \ref{fig0}, the 2D B-scan image represents a single slice of 3D volume, translating OCT B-scan to OCTA B-scan disregards the continuity across images. On the other hand, the 2D projection map is obtained through dimensional reduction, which may result in the loss of complete information present in 3D volume. As a result, learning with one type of images may lead to low quality in the other type. Secondly, current OCTA translation methods do not consider vessel information, which leads to poor quality in vascular areas of translated OCTA projection maps. This problem arises from the use of pixel-wise reconstruction loss functions that do not differentiate between vessel and non-vessel pixels. As OCTA images predominantly consist of non-vessel areas, the model tends to prioritize accuracy in non-vessel regions over vascular areas.

To address the above limitations, we propose a novel TransPro method with two main \mia{ideas}.
The first \mia{idea} is derived from a critical observation that the OCTA projection map is generated by averaging pixel values from its corresponding B-scans along the Z-axis. Hence, it is crucial to maintain this property by leveraging both B-scan and projection maps of 3D OCT. To achieve this, we propose a hybrid network architecture, consisting of a 3D generative network and a 2D generative network. \textbf{\textit{The 3D generative network takes \mia{3D OCT volume} as input}} and generates \mia{3D OCTA volume} as output.
During training, the 3D generative network is optimized by minimizing the disparities between the generated OCTA images and the ground-truth, both along the \mia{3D volumes} and the projection maps. Concurrently, \textbf{\textit{the 2D generative network takes OCT projection maps as inputs}} and generates OCTA projection maps as outputs.
To leverage the additional information acquired by the 2D network from OCT projection maps, we introduce a novel \textbf{heuristic contextual guidance (HCG)}, which minimizes the discrepancies between the outputs of the 2D network and the projection maps generated by the 3D network. By doing so, it serves as an additional mechanism to ensure the consistency of the generated OCTA projection maps between the 2D and 3D networks.

The second \mia{idea} aims to enhance the vessel area during the generation process to prevent excessive attention on the background regions. To achieve this, we introduce a \textbf{Vessel Promoted Guidance (VPG)}. By utilizing a pre-trained 2D retinal vascular segmentation model to segment vessels in 2D OCTA projection maps, our proposed VPG module ensures semantic consistency between the vascular segmentation predictions of the translated OCTA images and the ground-truth images.
By using these two novel approaches, our TransPro method achieves enhanced consistency between B-scans and their corresponding OCTA projection maps, as well as improved quality of blood vessels in the translated OCTA images.

We evaluate the proposed method on the two subsets of OCTA-500 public dataset~\cite{li2020ipn}. We compute Mean Absolute Error (MAE), Peak Signal-to-Noise Ratio (PSNR), and Structural Similarity Index (SSIM) on both translated OCTA B-scan images and projection maps. Moreover, we propose two novel metrics, named Vessel Density Error (VDE) and Vessel Density Correlation (VDC), to evaluate the quality of vessel regions in the translated OCTA projection maps. The experimental results demonstrate the superiority of the proposed approach over state-of-the-art methods. Our method achieves a remarkable 11.4\% relative improvement in the Mean Absolute Error (MAE) metric on the OCTA-3M daatset. Similarly, on the larger OCTA-6M dataset, the MAE was reduced by 5.1\% compared to prior art. Furthermore, our approach also yields substantial enhancements in vessel quality metrics. On the OCTA-3M dataset, there is a 40.8\% relative improvement in VDE and a 14.0\% boost in VDC. The OCTA-6M results are also impressive, with a 21.6\% VDE increase and an 8.6\% VDC increase.

The main contributions of this paper are summed up as follows: 
\begin{itemize}
    \item We identify a novel motivation in OCT-OCTA translation, which is to maintain the consistency of the translated OCTA images across both 3D volumes and projection maps. 

    \item We introduce a novel approach called Heuristic Contextual Guidance (HCG) to provide additional guidance aimed at preserving the quality of OCTA projection maps.
    
    \item We propose a novel Vessel Promoted Guidance (VPG) module, which significantly enhances the accuracy and quality of vessel regions in translated OCTA projection maps. 
    
    \item We evaluate our TransPro on translated OCTA B-scan images and projection maps. In addition to the commonly used metrics, we propose several vessel-specific metrics to quantitatively evaluate the vessel quality of translated OCTA in projection maps.
   
    \item Extensive experimental results demonstrate that our TransPro significantly outperforms state-of-the-art techniques across a range of performance criteria on both OCTA-3M and OCTA-6M datasets, including both traditional image translation metrics and clinically relevant vessel quality metrics. These results highlight the effectiveness and advantages of our proposed TransPro approach.
    
\end{itemize}
\section{Related Work}
\label{2:relatedwork}
\subsection{OCT and OCTA Image Analysis}
OCT and OCTA are widely used imaging modalities for diagnosing various \mia{retinal} diseases~\cite{huang1991optical,fujimoto2016development}. OCT enables non-invasive real-time imaging of the three-dimensional cross-sectional structure of the retina, but it does not provide sufficient information for detecting choroidal neovascularization or recognizing the morphological characteristics of microvasculature, which are important for diagnosing retinal diseases~\cite{spaide2018optical}. As a complementary modality to OCT, OCTA is introduced to visualize the retinal blood vessels~\cite{roisman2017oct,eladawi2018early}. OCTA images can be generated by detecting temporal signal changes caused by moving red blood cells in multiple OCT images obtained from the same location~\cite{ferrara2016investigating,kashani2017optical}. 
However, OCTA devices \mia{require additional sensors and software to alleviate the eye motion artifacts. The higher cost limits their wider usage}. 

\subsection{Modality Translation in Medical Images}
In clinical diagnosis, the utilization of multi-modality images can provide comprehensive and complementary information that can be advantageous for treatment~\cite{zhang2021texture,li2020deep}. However, the acquisition of multi-modality images poses challenges due to factors such as cost, time, and potential harm to the body~\cite{zhang2022deep}. To overcome this issue, a promising solution is to perform modality translation from an easily obtainable source image modality to a more challenging target image modality~\cite{mcnaughton2023machine}. Examples include translating magnetic resonance imaging (MRI) to computerized tomography (CT)~\cite{hsu2022synthetic,parrella2023synthetic,zhao2023saru}, MRI to positron emission tomography (PET)~\cite{rajagopal2022synthetic,hussein2022multi}, and PET to CT~\cite{li2022eliminating}. Among these, the translation of OCT to OCTA images has been an active area of research in recent years.

The existing OCT to OCTA translation methods can be divided into two groups according to the types of input images. The first group of methods translates OCT B-scan images to their corresponding OCTA B-scan images. For example, the initial work~\cite{lee2019generating} employs an encoder-decoder model structure to generate 2D OCTA B-scan images from paired 2D OCT B-scan images. Building upon this structure, Texture-UNet~\cite{zhang2021texture} extracts the texture features of OCT B-scans to guide the translation process. 
Recently, some methods utilize Generative Adversarial Network (GAN) to achieve improved performance in the OCT to OCTA translation task. AdjacentGAN~\cite{li2020deep} employs three adjacent OCT B-scan images as input to guide the translation of the middle OCTA B-scan image. This approach utilizes limited contextual information from neighboring OCT slices to enhance the continuity among slices.
The second group of methods focuses on the translation of 2D projection maps.
MultiGAN~\cite{pan2022multigan} proposes multiple GANs to translate OCT projection maps to OCTA projection maps obtained between different retinal layers.
However, the current methods only consider either B-scan images or projection maps for OCT to OCTA translation, which results in incomplete information about 3D volumes. This omission ignores the relevance between the two types of OCT and OCTA data, leading to low-quality translations in one type when training with the other one.
Moreover, the quality of vascular areas in translated OCTA projection maps is relatively poor due to the lack of specific vessel enhancement guidance.
To address these issues, this paper introduces a 3D GAN with an HCG module to constrain the translation of hybrid data formats, and a VPG module to improve the quality of vessel areas.

\subsection{Retinal Vascular Segmentation}
Retinal vascular segmentation of OCTA images is a crucial and fundamental task in the diagnosis of retinal diseases. The segmentation provides information on the density and morphological structures of retinal capillary networks, which can be utilized for disease diagnosis or robotic surgery~\cite{ronneberger2015u}. Various methods have been proposed for retinal vascular segmentation based on the classical UNet framework in recent years~\cite{ma2020rose,mou2019cs,li2020ipn}. For instance, CSNet adds self-attention blocks to the encoder and decoder network to emphasize curvilinear vessel structures~\cite{mou2019cs}. Besides, DCSSNet is designed as a semi-supervised segmentation model that uses both labeled and unlabeled images to mitigate the high cost of pixel-level annotations~\cite{chen2022dual}. Existing methods consider retinal vascular segmentation as an independent medical image analysis task and treat model design and downstream tasks separately. In contrast, our work treats OCTA vascular segmentation as an auxiliary task to enhance the quality of vascular regions in the translated OCTA images, rather than solely solving the vascular segmentation problem. This approach provides a novel perspective and demonstrates the potential of OCTA vascular segmentation as an auxiliary task.

\subsection{Multi-Task Learning}
Multi-task learning is a widely used approach that aims to optimize a machine learning model for multiple tasks in a unified framework~\cite{caruana1997multitask}. It employs shared feature representations to achieve balanced optimization for all the tasks~\cite{li2021jigsawgan}. Despite differences in the requirements for image features, combining adjacent tasks such as object detection and instance segmentation has shown remarkable empirical results~\cite{zhang2020feature,zhang2021self}. In cases where tasks have unequal importance, auxiliary learning, where auxiliary tasks are used to assist the main tasks in achieving better predictions, can be employed~\cite{kendall2018multi,liu2019end}. Auxiliary tasks are frequently used in generative adversarial networks, such as classification and jigsaw solving~\cite{liebel2018auxiliary,odena2017conditional}. In our method, we use two auxiliary tasks, vascular segmentation, and 2D OCTA projection map translation, to help the main task, OCT to OCTA image translation, learn rich feature representations. We integrate all tasks into the same framework and train them in an end-to-end fashion. Our contribution is using two auxiliary tasks to improve the performance of the main task. The coherence between the auxiliary tasks and the main task in terms of features enhances the performance of OCT to OCTA image translation.

\begin{figure*}[t]
\includegraphics[width=.95\textwidth]{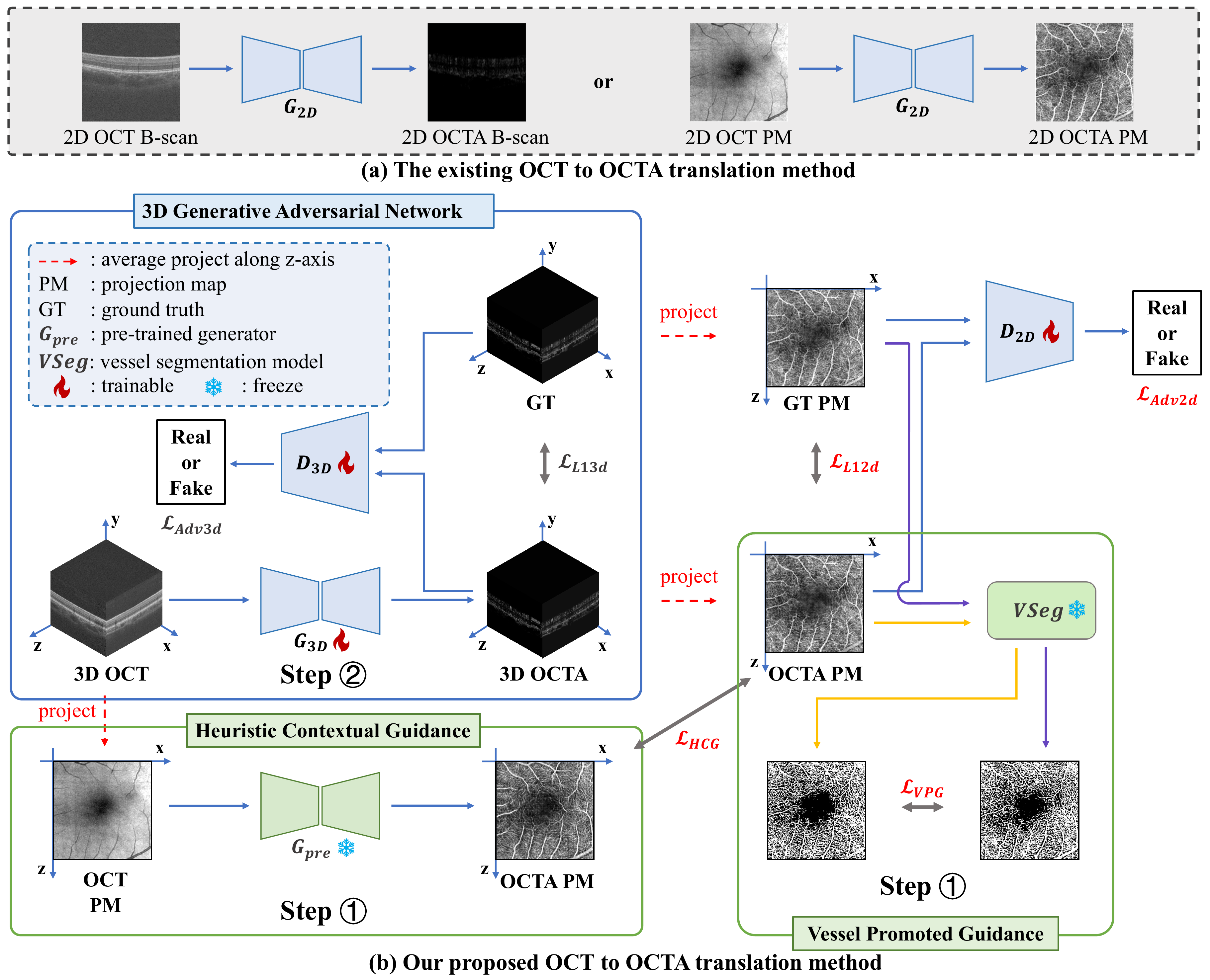}
\centering
\vspace{-2mm}
\caption{The main ideas and contributions of this work. (a) illustrates the overall framework of the existing OCT to OCTA translation methods. (b) illustrates our proposed TransPro. Compared to existing methods, TransPro not only takes into considerations of critical vessel information via the Vessel-Promoted Guidance model but also introduces beneficial contexts via the Heuristic Contextual Guidance model.} 
\vspace{-2mm}
\label{fig2}
\end{figure*}
\section{Methodology}
\label{3:ours}
As illustrated in Figure \ref{fig2} (b), the proposed TransPro mainly contains the following three components: 1) a 3D conditional generative adversarial network in \mia{Section}~\ref{part1}, which is the main body of our method, 2) a Heuristic Contextual Guidance (HCG) module in \mia{Section}~\ref{part2} to align the translation of projection maps.
3) a Vessel Promoted Guidance (VPG) module in \mia{Section}~\ref{part3} to improve the quality of vessel areas. \mia{We summarize the entire TransPro framework in Section~\ref{overall}.}

\subsection{3D Generative Adversarial Network}
\label{part1}
OCT to OCTA image translation task aims to translate an OCT volume $\textbf{X} \in \mathbb{R}^\mia{{L \times W \times D}}$ to its paired OCTA volume $\hat{\textbf{Y}} \in \mathbb{R}^\mia{{L \times W \times D}}$, where the translated OCTA volume $\hat{\textbf{Y}}$ needs to be as close as possible to the ground-truth OCTA volume $\textbf{Y} \in \mathbb{R}^\mia{{L \times W \times D}}$~\cite{yang2023deep}. In particular, each volume data is composed of a set of 2D slices with the size of \mia{$L \times D$} (\ie, B-scan images) and the number of slices for each volume is \mia{$W$}. 

Our proposed framework builds upon the classical image-to-image translation model, pix2pix~\cite{isola2017image}. Unlike existing methods that only consider the OCT B-scan image as input to the generator to produce the corresponding OCTA B-scan image~\cite{roisman2017oct,eladawi2018early}, we extend the network to three-dimensional convolutional neural networks to handle 3D volumes and enhance the global contextual learning across slices. Specifically, the input to the 3D generator is a 3D OCT volume and the output is a translated OCTA volume. Additionally, a 3D discriminator is used to differentiate between ground-truth and translated OCTA volumes. The 3D generator and discriminator are denoted as $G_{3d}$ and $D_{3d}$, respectively, and the adversarial loss is formulated accordingly:
\begin{equation}
\begin{aligned}
\mathcal{L}_{Adv3d}= &~\mathbb{E}_{\textbf{Y}\sim p(OCTA)}[\log(D_{3d}(\textbf{Y}))] + \\
&~\mathbb{E}_{\textbf{X}\sim p(OCT)}[\log(1-(D_{3d}(G_{3d}(\textbf{X}))))], \label{Adv3d_loss} 
\end{aligned}
\end{equation}
where $\textbf{Y}$ denotes the ground-truth OCTA volume sampled from distribution $p(OCTA)$ and $\textbf{X}$ denotes the input OCT volume sampled from distribution $p(OCT)$. $G_{3d}(\textbf{X})$ which is equal to $\hat{\textbf{Y}}$, denotes the output OCTA volume translated by $G_{3d}$ network.
In addition, there is a constraint on each voxel between the translated volume $G_{3d}(\textbf{X})$ and the ground-truth volume $\textbf{Y}$. To ensure less blurring, we follow \cite{isola2017image} and apply $L1$ distance instead of $L2$, which is formulated as:
\begin{equation}
\begin{aligned}
\mathcal{L}_{L13d}=~\left\|\textbf{Y}-G_{3d}(\textbf{X})\right\|_1.
\label{L13d_loss} 
\end{aligned}
\end{equation}

In addition to the 3D volume \mia{$(L \times W \times D)$}, there is another type of data format, the 2D projection map \mia{$(L \times W)$} data that is capable of visualizing vascular structures. The projection maps are obtained by averaging the values of pixels along the \mia{$D$} dimension. We denote the projection process as $Proj(\cdot)$, and the obtained OCTA projection map as $\hat{\textbf{y}}\in \mathbb{R}^\mia{{L \times W}}$, then we have: 
\begin{equation}
\begin{aligned}
\hat{\textbf{y}} = Proj(\hat{\textbf{Y}}).\label{projection} 
\end{aligned}
\end{equation}
Based on experimental observations, it has been noted that the constraints in Equation~\mia{\ref{Adv3d_loss} and~\ref{L13d_loss}} may not be sufficient to ensure high quality in the translated OCTA projection maps. The reason behind this is the accumulation of errors during the computation of projection maps of the translated OCTA volumes with Equation~\ref{projection}. Therefore, to enhance the similarity between the ground-truth and translated OCTA projection maps, we propose the use of a 2D adversarial loss (referred to as $\mathcal{L}_{Adv2d}$ \mia{in Equation~\ref{Adv2d_loss}}) and L1 loss (referred to as $\mathcal{L}_{L12d}$ \mia{in Equation~\ref{projection_loss}}). To achieve this, we apply the projection function $Proj(\cdot)$ on the ground-truth OCTA volume $\textbf{Y}$ and the translated OCTA volume $\hat{\textbf{Y}}$ to obtain the corresponding ground-truth and translated projection maps, denoted as $\textbf{y}$ and $\hat{\textbf{y}}$, respectively. The losses are then computed as follows:

\begin{equation}
\begin{aligned}
\mathcal{L}_{Adv2d}= &~\mathbb{E}_{\textbf{Y}\sim p(OCTA)}[\log(D_{2d}(\textbf{y}))]+ \\
&~\mathbb{E}_{\textbf{X}\sim p(OCT)}[\log(1-(D_{2d}(\hat{\textbf{y}})))],\label{Adv2d_loss} 
\end{aligned}
\end{equation}
\begin{equation}
\begin{aligned}
\mathcal{L}_{L12d}(G_{3d})
~=&~\left\|~\textbf{y}-\hat{\textbf{y}}~\right\|_1,
\label{projection_loss} 
\end{aligned}
\end{equation}
where 
\begin{equation}
\begin{aligned}
\textbf{y} &= Proj(\textbf{Y}), \\
\hat{\textbf{y}} &= Proj(\hat{\textbf{Y}}) = Proj(G_{3d}(\textbf{X})).
\label{explaination1} 
\end{aligned}
\end{equation}
$D_{2d}$ denotes a 2D discriminator which differentiates the ground-truth and translated OCTA projection maps.

The overall loss function for the 3D generative adversarial network is expressed as:
\begin{equation}
\begin{aligned}
\mathcal{L}_{3DGAN}= \mathcal{L}_{Adv3d} + \mathcal{L}_{Adv2d} + \mia{\lambda_1}\mathcal{L}_{L13d}+\mia{\lambda_2}\mathcal{L}_{L12d},
\label{3dGAN_loss} 
\end{aligned}
\end{equation}
where \mia{$\lambda_1$ and $\lambda_2$} are loss-balance hyperparameters which are set to 10 in our experiments. \mia{We set them to an equal value to keep the hyperparameter search computationally feasible.} 

\subsection{Heuristic Contextual Guidance (HCG)}
\label{part2}
\mia{The loss function described in Equation \ref{projection_loss} enforces the alignment between the translated and the real OCTA images in the view of projection maps. However, we observe that there exist some vessel discontinuity regions in real OCTA projection maps due to the unstable scanning by OCTA devices. Therefore, the model may suffer the overfitting problem by learning these specific patterns. To alleviate this issue, we introduce the Heuristic Contextual Guidance (HCG) module, which is a pre-trained OCT to OCTA projection map translation model. By leveraging the convolution operations applied to 2D OCTA projection maps in the HCG module, each output pixel contains information from its neighboring pixels. This spatial context plays a crucial role in improving vessel continuity in 2D OCTA projection maps.}

Specifically, the pre-trained model's generator ($G_{pre}$) is utilized and its parameters are fixed during training. The projection map of the input OCT volume is then subjected to $G_{pre}$, resulting in an OCTA projection map that contains independent contextual information from the projection view. The output of the 2D translation model, $G_{pre}$, is denoted as $\textbf{y}^{\prime}$. The $L1$ distance is utilized to determine the dissimilarity between $\textbf{y}^{\prime}$ and the projection map of translated OCTA, $G_{3d}(\textbf{X})$, which is produced by $G_{3d}$, which can be formulated as:
\begin{equation}
\begin{aligned}
\mathcal{L}_{HCG}(G_{3d})
~=&~\left \|~\textbf{y}^{\prime}-Proj(G_{3d}(\textbf{X}))~\right\|_1,
\label{context_loss} 
\end{aligned}
\end{equation}
where 
\begin{equation}
\begin{aligned}
\textbf{y}^{\prime}~=&~G_{pre}(Proj(\textbf{X})).
\label{explaination3} 
\end{aligned}
\end{equation}

\subsection{Vessel Promoted Guidance (VPG)}
\label{part3}
Given that OCTA images are primarily used to reflect vessel signals, the similarity of blood vessels between synthesized and ground-truth data is more crucial than the similarity of background tissues. However, it is challenging for the network to focus on vessel pixels in the absence of specific flow information~\cite{ferrara2016investigating,kashani2017optical}. Thus, we introduce a vessel segmentation model with fixed parameters to guide the generation accuracy of vessel regions. 

Moreover, during the 3D volume translation process, the lack of vessel information leads to aimlessness, resulting in poor quality of vascular regions in translated OCTA projection maps. To this end, we propose a vascular segmentation model to focus more on the vascular regions. Initially, we pre-train a vascular segmentation model $VSeg(\cdot)$ on OCTA projection maps with annotated vessel labels. Next, we use $VSeg(\cdot)$ to extract semantic vascular segmentation logits on both ground-truth and translated OCTA projection maps, denoted as $\textbf{l}_{seg}$ and $\hat{\textbf{l}}_{seg}$, respectively. Finally, we employ L1 distance to minimize the discrepancy between the two logits:
\begin{equation}
\begin{aligned}
\mathcal{L}_{VPG}~=&~\left \|~\hat{\textbf{l}}_{seg}- \textbf{l}_{seg}~\right\|_1, 
\label{vesselseg_loss} 
\end{aligned}
\end{equation}
where 
\begin{equation}
\begin{aligned}
\hat{\textbf{l}}_{seg}~=&~VSeg(Proj(G_{3d}(\textbf{X}))), \\
\textbf{l}_{seg}~=&~VSeg(Proj(\textbf{Y})).
\label{explaination2} 
\end{aligned}
\end{equation}

The $\mathcal{L}_{VPG}$ loss \mia{(Equation~\ref{vesselseg_loss})} aims to match the vascular structures in the ground-truth and translated OCTA projection maps, addressing the problem of aimlessness in vascular areas.

\subsection{Overall Framework}
\label{overall}
The overall framework of our proposed TransPro method consists of three parts, which are 3D GAN, the HCG module, and the VPG module. We pre-train the models for VPG and HCG modules and fix their parameters when training the 3D GAN.

During training, we jointly train the generator $G_{3d}$ and two discriminators $D_{3d}$ and $D_{2d}$ following the generative-adversarial learning pattern~\cite{isola2017image,zhu2017unpaired}. Our total loss function can be formulated as:
\begin{equation}
\begin{aligned}
\mathcal{L}_{TransPro} = \mathcal{L}_{3DGAN}
+ \mia{\alpha}\mathcal{L}_{VPG} + \mia{\beta}\mathcal{L}_{HCG},
\label{total_loss} 
\end{aligned}
\end{equation}
where \mia{$\alpha$ and $\beta$} are task-balance hyperparameters which are set to 5 in our experiments. \mia{We set the weights to an equal value to keep the hyperparameter search computationally feasible.} 
The final objective function is expressed as:
\begin{equation}
\begin{aligned}
&G^*_{3d} = \arg\mathop{\min}\limits_{G_{3d}}\mathop{\max}\limits_{D_{3d},D_{2d}} \mathcal{L}_{TransPro}(G_{3d},D_{3d},D_{2d}).
\label{total_object} 
\end{aligned}
\end{equation}
In inference, only the 3D generator $G_{3d}$ is used. Therefore, TransPro does not require any additional computational overheads in this process. 
\section{Experiments}
\label{4:exps}
\begin{figure*}[t]
\includegraphics[width=.98\textwidth]{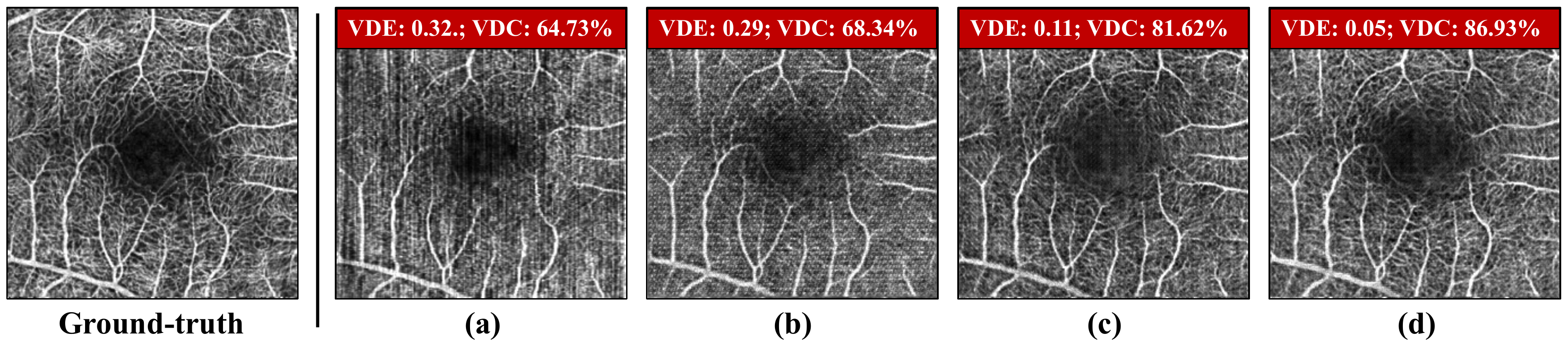}
\centering
\vspace{-2mm}
\caption{The OCTA projection maps with different vessel density error (VDE) and vessel density correlation (VDC) results. The leftmost side is the ground-truth and (a)-(d) are the four translated OCTA projection maps generated by different OCT to OCTA translation methods with different qualities from low to high, \mia{namely Adjacent GAN~\cite{li2020deep}, Pix2pix 3D~\cite{isola2017image}, TransPro (not the best epoch), and TransPro (the best epoch), respectively.} The results of VDE and VDC metrics are matched with the visualized quality.}
\label{fig4}
\end{figure*}
\subsection{Dataset (OCTA-500)}
OCTA-500~\cite{li2020ipn} is a publicly accessible dataset with 500 pairs of 3D OCT and OCTA volumes. This dataset is divided into two subsets according to the field of view types, which are OCTA-3M and OCTA-6M, respectively. We implement our model on these two subsets.
\begin{itemize}
\item \textbf{OCTA-3M} contains 200 pairs of OCT and OCTA volumes with \mia{3mm $\times$ 3mm $\times$ 2mm} field of view. Each volume is with the size of \mia{304px $\times$ 304px $\times$ 640px}, and the size of the projection map is 304px $\times$ 304px. Following~\cite{li2020ipn}, we divide the size of the training set, validation set, and test set as 140 volumes, 10 volumes, and 50 volumes, respectively.
\item \textbf{OCTA-6M} contains 300 pairs of OCT and OCTA volumes with \mia{6mm $\times$ 6mm $\times$ 2mm} field of view. The volume is with the size of \mia{400px $\times$ 400px $\times$ 640px}, and the size of its projection map is 400px $\times$ 400px. The size of the training set, validation set, and test set for the OCTA-6M dataset is 180 volumes, 20 volumes, and 100 volumes, respectively.
\end{itemize}
 
\subsection{Evaluation Metrics}
We evaluate the translated OCTA images from the following two aspects: OCTA volume and OCTA projection map.
To evaluate the quality of OCTA volume, we use standard image quality assessment metrics, \ie, Mean Absolute Error (\textbf{MAE}), Peak Signal-to-Noise Ratio (\textbf{PSNR})~\cite{qpsnr}, and Structural SIMilarity (\textbf{SSIM})~\cite{wang2004image} and average the results over each slide of the whole volume (B-scan image).

In addition to the standard evaluation metrics used to assess the quality of OCTA volume, we implement five supplementary evaluation metrics specifically designed for the OCTA projection maps. \mia{These metrics serve to quantify the similarity of vessel topology and morphology structures in OCTA projection maps, which are routinely examined by doctors for disease diagnosis}.

The first three metrics are modified from the MAE, PSNR, and SSIM metrics and are now referred to as vessel-weighted metrics. We denote these metrics as \textbf{MAE-V}, \textbf{PSNR-V}, and \textbf{SSIM-V}.
We annotate binary pixel-wise vessel segmentation mask for each ground-truth OCTA projection map, denoted as $\hat{\textbf{M}}$, that 0 represents non-vessel pixels and 1 represents vessel pixels.
We preprocess the translated and the ground-truth OCTA projection maps to decrease the weights of non-vessel pixels. The values of non-vessel pixels are multiplied by a ratio $\gamma \in (0, 1.0]$, while the vessel pixels values remain unchanged. We denote this preprocessing process as $\textbf{V}(\cdot)$ that is defined as:

\begin{equation}
\label{weight}
\textbf{V}(\textbf{I}(i,j))=\left\{
\begin{aligned}
\gamma * \textbf{I}(i,j) & , & if~\hat{\textbf{M}}(i,j) = 0 \\
\textbf{I}(i,j) & , & if~\hat{\textbf{M}}(i,j) = 1
\end{aligned}
\right.,
\end{equation}
where $(i,j)$ denotes the position of pixels.

Then, the vessel-weighted metrics are calculated as:

\begin{equation}
\begin{aligned}
\textbf{MAE\text{-}V}=\frac{1}{N}\sum_{i=1}^N \textbf{MAE}(\textbf{V}(\textbf{I}), \textbf{V}(\hat{\textbf{I}})),
\label{mae-v} 
\end{aligned}
\end{equation}

\begin{equation}
\begin{aligned}
\textbf{PSNR\text{-}V}=\frac{1}{N}\sum_{i=1}^N \textbf{PSNR}(\textbf{V}(\textbf{I}), \textbf{V}(\hat{\textbf{I}})),
\label{psnr-v} 
\end{aligned}
\end{equation}

\begin{equation}
\begin{aligned}
\textbf{SSIM\text{-}V}=\frac{1}{N}\sum_{i=1}^N \textbf{SSIM}(\textbf{V}(\textbf{I}), \textbf{V}(\hat{\textbf{I}})),
\label{ssim-v} 
\end{aligned}
\end{equation}
where $\textbf{I}$ represents the translated OCTA projection maps and $\hat{\textbf{I}}$ denotes the ground-truth. $N$ is the size of test set.

We set the value of $\gamma$ to 0.1 for the results presented in Table~\ref{table2}. Additionally, we provide results for the MAE-V, PSNR-V, and SSIM-V metrics for other values of $\gamma$ in Section~\ref{Ablation-2}.

In addition, we propose two novel metrics based on vessel density~\cite{yao2020quantitative}, \ie, \textbf{V}essel \textbf{D}ensity \textbf{E}rror (\textbf{VDE}) and \textbf{V}essel \textbf{D}ensity \textbf{C}orrelation (\textbf{VDC}). Vessel density is a crucial clinical measure used to detect vessel abnormalities in various retinal diseases, particularly in their early phases~\cite{mastropasqua2017foveal,al2018biomarkers,richter2018diagnostic}. When comparing two OCTA projection maps, if they possess the same vessel density, they can be considered identical. The definition of vessel density is the proportion of the vessel area over the total area~\cite{alam2017computer}, which can be expressed as the percentage of vessel pixels over the total pixels within a given region.
\mia{We adopt a traditional image processing approach as described in~\cite{levine2020repeatability} for obtaining vessel segmentation masks. This method involves binarizing the OCTA projection maps using a global mean threshold, where pixels above the threshold are interpreted as vessels.}
We denote the obtained binary masks as $\textbf{M}$ and $\hat{\textbf{M}}$, for translated and ground-truth OCTA projection maps, respectively. Then, the vessel density is calculated as the ratio of vessel area to the entire imaged area.

\begin{table*} [t]
\centering
\vspace{-2mm}
\caption{Result comparisons with the state-of-the-arts on OCTA volumes and OCTA projection maps of OCTA-3M dataset~\cite{li2020ipn}. The MAE, PSNR (dB), and SSIM (\%) are computed on OCTA volumes, while the MAE-V, PSNR-V (dB), SSIM-V (\%), VDE, and VDC (\%) are computed on OCTA projection maps. Results of other methods are produced by our re-implementation. ``$\downarrow$'' refers to the lower the better and ``$\uparrow$'' refers to the higher the better.}
\resizebox{\linewidth}{!}{
\begin{tabular}{r|ccc|ccccc}
\toprule
Method & MAE~$\downarrow$ & PSNR(dB)~$\uparrow$ & SSIM(\%)~$\uparrow$ & MAE-V~$\downarrow$ & PSNR-V(dB)~$\uparrow$ & SSIM-V(\%)~$\uparrow$ & VDE$\downarrow$ & VDC(\%)$\uparrow$\\
\midrule
Pix2pix 2D~\cite{isola2017image} &  0.0968 &  29.59 & 84.59 & 0.0823 & 17.92 & 85.26 & \mia{0.1764} & \mia{62.49} \\
Pix2pix 3D~\cite{isola2017image} & 0.0883 & 31.58 & 86.24 & 0.0753 & 20.36 & 89.14 & \mia{0.2206} & \mia{65.29} \\
9B18CN UNet~\cite{lee2019generating} & 0.0918 & 29.34 & 86.04 & 0.0839 & 17.12 & 85.03 & \mia{0.1612} & \mia{63.52} \\
Adjacent GAN~\cite{li2020deep} & 0.0906 & 30.89 & 87.14 & 0.0833 & 17.56 & 85.88 & \mia{0.2034} & \mia{63.05} \\
VQ-I2I~\cite{chen2022eccv} & 0.0824 & 31.72 & 87.59 & 0.0760 & 18.61 & 88.90 & \mia{0.1469} & \mia{71.07}\\
Palette~\cite{saharia2022palette} & 0.0814 & 32.42 & \textbf{88.24} & 0.0793 & 19.17 & 87.97 & \mia{0.1466} & \mia{70.86}\\
\textbf{TransPro (ours)} & \textbf{0.0782} & \textbf{32.56} & {88.22} & \textbf{0.0658} & \textbf{20.42} & \textbf{91.79} & \textbf{\mia{0.1304}} & \textbf{\mia{74.41}}\\
\bottomrule
\end{tabular}}
\label{table1}
\end{table*}

\begin{itemize}
\item \textbf{Vessel Density Error (VDE).} The VDE metric is utilized to assess the absolute difference of vessel density between translated and ground-truth OCTA projection maps. It is computed as 
\begin{equation}
\begin{aligned}
\textbf{VDE}=\frac{1}{N}\sum_{i=1}^N | \textbf{M}_{VD} - \hat{\textbf{M}}_{VD}| ,
\label{vde} 
\end{aligned}
\end{equation}
where N is the size of test set and the subscript VD denotes the vessel density of the image.
As shown in Figure~\ref{fig4}, lower VDE values indicate better quality of the translation outcomes.
\item \textbf{Vessel Density Correlation (VDC).} VDC is employed to assess the similarity of vessel density in a local-wise manner. Specifically, we partition each OCTA projection map into small patches $\textbf{m}^i$ of size 16px $\times$ 16px, compute the vessel density for each patch, and derive an array of regional vessel densities, denoted as $\textbf{M}_{VD}^{arr}$. 

\begin{equation}
\begin{aligned}
\textbf{M}_{VD}^{arr}= [\textbf{m}_{VD}^1, \textbf{m}_{VD}^2, ..., \textbf{m}_{VD}^k],
\label{array} 
\end{aligned}
\end{equation}

The VDC is calculated as the Pearson correlation coefficient of the two arrays between the translated and ground-truth OCTA projection maps.

\begin{equation}
\begin{aligned}
\textbf{VDC} = \frac{1}{N}\sum_{i=1}^N\frac{cov(\textbf{M}_{VD}^{arr},\hat{\textbf{M}}_{VD}^{arr})}{\sigma_{\textbf{M}_{VD}^{arr}}\sigma_{\hat{\textbf{M}}_{VD}^{arr}}},
\label{vdc} 
\end{aligned}
\end{equation}
where N is the size of test set, $cov$ is the covariance, and $\sigma$ is the standard 
deviation. Examples in Figure~\ref{fig4} indicate that a higher VDC corresponds to a better translation quality.
\end{itemize}

\subsection{Implementation Details}

\begin{table*} [t]
\centering

\vspace{-2mm}
\caption{Result comparisons with the state-of-the-arts on OCTA volumes and OCTA projection maps of OCTA-6M dataset~\cite{li2020ipn}. The MAE, PSNR (dB), and SSIM (\%) are computed on OCTA volumes, while the MAE-V, PSNR-V (dB), SSIM-V (\%), VDE, and VDC (\%) are computed on OCTA projection maps. Results of other methods are produced by our re-implementation. ``$\downarrow$'' refers to the lower the better and ``$\uparrow$'' refers to the higher the better.}
\resizebox{\linewidth}{!}{
\begin{tabular}{r|ccc|ccccc}
\toprule
Method & MAE$\downarrow$ & PSNR(dB)$\uparrow$ & SSIM(\%)$\uparrow$ & MAE-V$\downarrow$ & PSNR-V(dB)$\uparrow$ & SSIM-V(\%)$\uparrow$ & VDE$\downarrow$ & VDC(\%)$\uparrow$ \\ 
\midrule
Pix2pix 2D~\cite{isola2017image} & 0.0995 & 27.65 & 87.15 & 0.0845 & 16.72 & 85.47 & \mia{0.1884} & \mia{63.33} \\
Pix2pix 3D~\cite{isola2017image} & 0.0900 & \textbf{30.66} & 87.16 & 0.0923 & 17.04 & 84.51 & \mia{0.1904} & \mia{67.68} \\
9B18CN UNet~\cite{lee2019generating} & 0.1135 & 27.91 & 83.69 & 0.0895 & 16.89 & 84.07 & \mia{0.2263} & \mia{62.96} \\
Adjacent GAN~\cite{li2020deep} & 0.1021 & 28.05 & 85.03 & 0.0879 & 17.19 & 84.44 & \mia{0.2052} & \mia{65.70} \\
VQ-I2I~\cite{chen2022eccv} & 0.0897 & 29.54 & 86.90 & 0.0788 & 17.38 & 86.78 & \mia{0.1523} & \mia{70.50} \\
Palette~\cite{saharia2022palette} & 0.0881 & 30.02 & 87.13 & 0.0797 & 17.95 & 85.78 & \mia{0.1575} & \mia{70.11}\\

\textbf{TransPro (ours)} & \textbf{0.0854} & 30.53 & \textbf{88.35} & \textbf{0.0733} & \textbf{18.30} & \textbf{90.23} & \textbf{\mia{0.1492}} & \textbf{\mia{73.47}}\\
\bottomrule
\end{tabular}}
\label{table2}
\end{table*}

\noindent\textbf{Settings.} The proposed TransPro model is trained in two steps. Firstly, we pre-train two models for VPG and HCG and freeze their parameters. Secondly, we train 3D GAN following the generative adversarial learning pattern with the total loss function (Equation~\ref{total_loss}). The backbones and the detailed training settings for each component are introduced below.

\begin{itemize}
\item \textbf{VPG.} We utilize the UNet~\cite{ronneberger2015u} as the underlying model for the vascular segmentation module in the VPG framework. This module aims to generate predictions of vascular segmentation in OCTA projection maps, using vessel pixel annotations as the ground-truth labels. Apart from the large vessel annotations provided in OCTA-500 dataset, we also annotate the pixel-wise labels for capillaries in OCTA projection maps. The annotations are obtained by two experts. The splitting of training, validation, and test sets is consistent with original 3D datasets.
We apply common data augmentation methods such as random cropping and flipping. The model is trained with cross-entropy loss using RMSprop optimizer with a learning rate of $10^{-5}$ for 100 epochs. After training, we choose the model with the best result on the validation set. 
\item \textbf{HCG.} We apply the pix2pix~\cite{isola2017image} as the backbone of the OCT to OCTA projection map translation model for HCG module. The projection maps of OCT and OCTA volumes are computed by the projection function $Proj(\cdot)$ in Equation~\ref{projection}. The splitting of training, validation, and test sets is consistent with original 3D datasets. We follow the settings in~\cite{isola2017image} and train the model over 200 epochs. After training, we select the model with the best result on the validation set. 
\item \textbf{TransPro.} Upon completion of training the VPG and HCG modules, we proceed with training the TransPro model. The 3D pix2pix model, a modified version of the model proposed in~\cite{isola2017image}, is used as the backbone of TransPro. In this setup, the 3D generator $G_{3d}$ and two discriminators $D_{3d}$ and $D_{2d}$ are jointly trained following the adversarial-learning manner. We use Adam optimizer and set the initial learning rate to 0.0002. The model is trained with a batch size of 1 for 200 epochs. To prevent overfitting, we evaluate the validation set using \mia{MAE} and select the best model with the minimum error value.
\end{itemize}
\begin{figure*}[t]
\includegraphics[width=.8\textwidth]{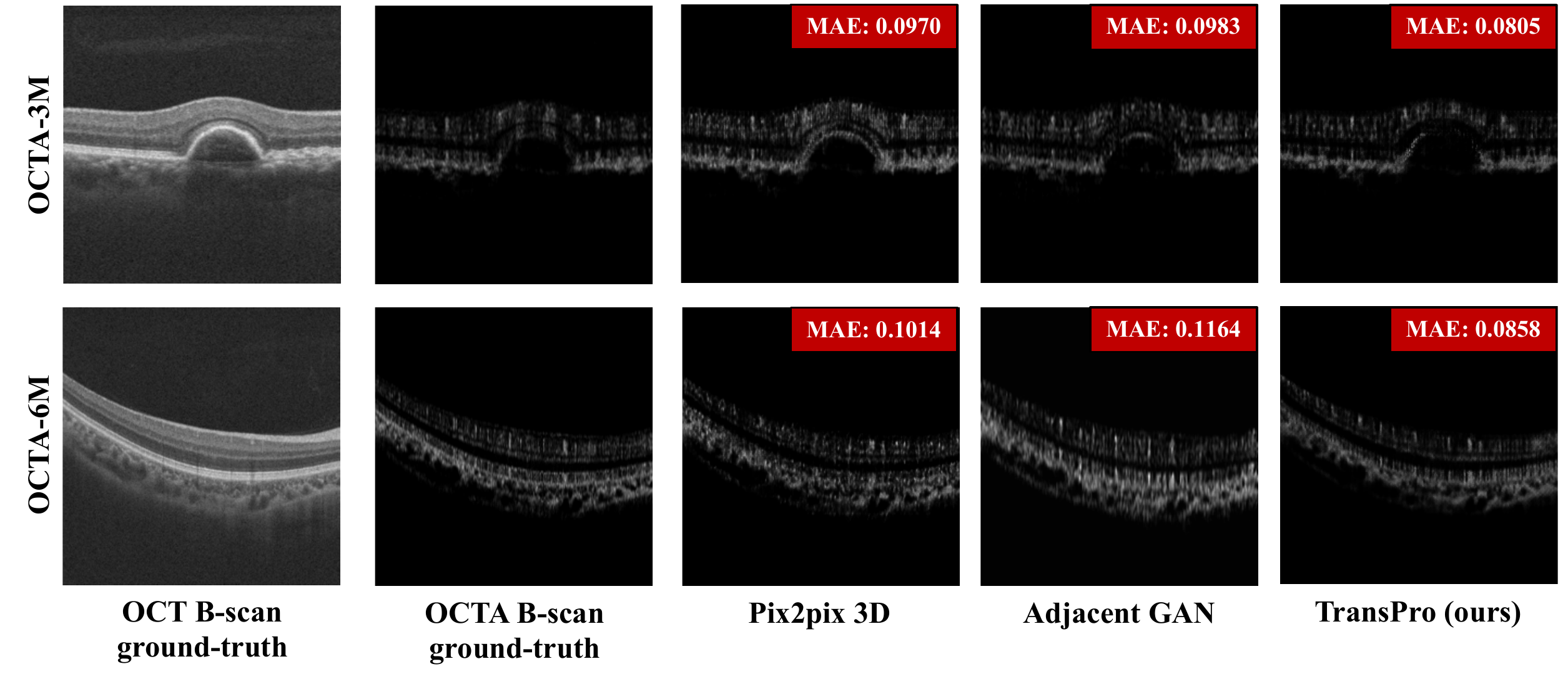}
\centering
\caption{The visualization examples of translated OCTA B-scans in OCTA-3M dataset and OCTA-6M dataset~\cite{li2020ipn}. We compared the results of different methods including pix2pix 3D, Adjacent GAN, and our proposed TransPro. The Mean Absolute Error (MAE) between the translated and the ground-truth OCTA B-scan images is computed.}
\label{fig3.0} 
\end{figure*}
\begin{figure*}[t]
\includegraphics[width=.99\textwidth]{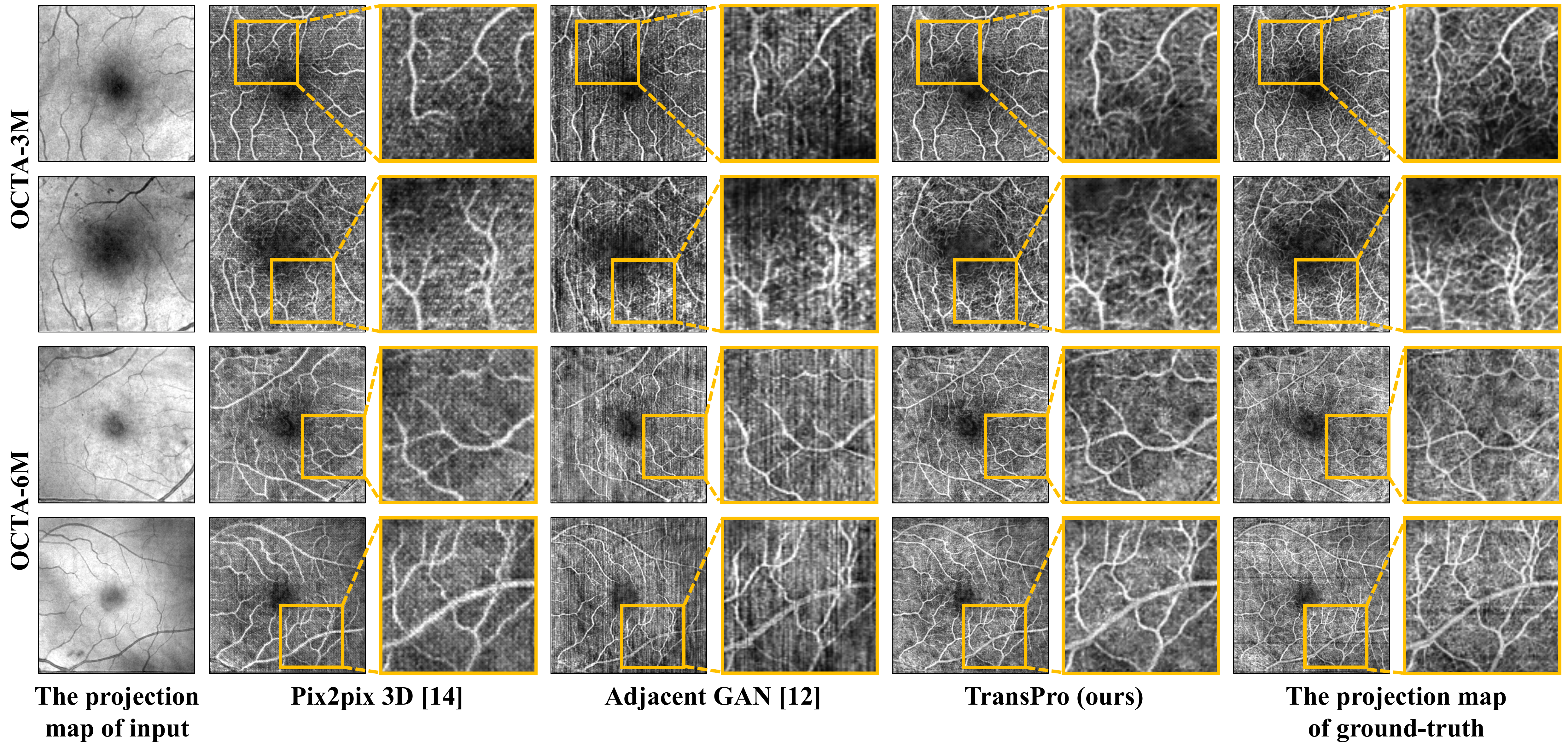}
\centering
\caption{Visualization result comparisons of translated OCTA projection maps in OCTA-3M dataset and OCTA-6M~\cite{li2020ipn}. Two examples are selected from each dataset correspondingly. We can observe that our approach can achieve better predictions. }
\label{fig3}
\end{figure*}
\begin{table*}
\centering
\setlength{\tabcolsep}{14pt}{
\caption{Ablation studies on the effectiveness of VPG and HCG modules on OCTA-3M and OCTA-6M~\cite{li2020ipn}. The MAE, PSNR (dB) and SSIM (\%) are computed on OCTA volumes, while the VDE and VDC (\%) are computed on OCTA projection maps. ``$\downarrow$'' refers to the lower the better and ``$\uparrow$'' refers to the higher the better.}
\begin{tabular}{ccc|ccc|cc}
\toprule
3D GAN & VPG & HCG & MAE~$\downarrow$ & PSNR(dB)~$\uparrow$ & SSIM(\%)~$\uparrow$ & VDE~$\downarrow$ & VDC(\%)~$\uparrow$\\
\midrule
& & & \multicolumn{5}{c}{OCTA-3M Dataset} \\
\hline
\checkmark & & & 0.0883 & 31.5824 & 86.24 & \mia{0.2206} & \mia{65.29} \\
\checkmark & \checkmark & & 0.0807 & 32.4794 & 87.43 & \mia{0.1791} & \mia{70.06} \\
\checkmark & & \checkmark & 0.0797 & 31.9813 & 87.78 & \mia{0.1682} & \mia{70.31}\\
\checkmark & \checkmark & \checkmark & \textbf{0.0782} & \textbf{32.5587}  & \textbf{88.22}  & \textbf{\mia{0.1304}} & \textbf{\mia{74.41}} \\
\hline
& & & \multicolumn{5}{c}{OCTA-6M Dataset} \\
\hline
\checkmark & & & 0.0900 & \textbf{30.6621} & 87.16 & \mia{0.1904} & \mia{67.68} \\
\checkmark & \checkmark & & 0.0879 & 30.1424 & 88.15 & \mia{0.1739} & \mia{71.04}\\
\checkmark & & \checkmark & 0.0860 & 29.8427 & 88.24 & \mia{0.1770} & \mia{69.56}\\
\checkmark & \checkmark & \checkmark & \textbf{0.0854} & {30.5264}  & \textbf{88.35}  & \textbf{\mia{0.1492}} & \textbf{\mia{73.47}} \\
\bottomrule
\end{tabular}
\label{table3}}
\end{table*}
 
\subsection{Comparisons with the State-of-the-art Methods}
We re-implement several state-of-the-art methods for OCT to OCTA image translation. The results of the comparisons of our proposed TransPro and other methods on OCTA-3M and OCTA-6M datasets are summarized in Table~\ref{table1} and~\ref{table2}, respectively. For each OCTA subject, we present the results of three commonly used metrics, MAE, PSNR, and SSIM, applied to the OCTA volume. Additionally, we introduce five novel vessel-specific metrics, namely MAE-V, PSNR-V, SSIM-V, Vessel Density Error (VDE), and Vessel Density Correlation (VDC), calculated on OCTA projection map. These proposed metrics provide a comprehensive assessment of both overall OCTA volume quality and vessel-specific characteristics, enabling a more detailed evaluation of the translated results.

\subsubsection{Experimental results on OCTA-3M} 
In our comparative experiments, We evaluate our TransPro method against two baseline methods, two OCT to OCTA image translation methods, and two state-of-the-art image-to-image translation methods. The re-implemented results for OCTA-3M dataset are displayed in Table~\ref{table1}.
The two baseline methods, namely pix2pix 2D and pix2pix 3D, are implemented following the network architectures described in \cite{isola2017image}. Pix2pix 2D takes paired OCT and OCTA B-scan images as inputs and outputs, while pix2pix 3D operates on paired OCT and OCTA volumes.
Additionally, we compare our method with Adjacent GAN \cite{li2020deep} and 9B18CN UNet \cite{lee2019generating}, which are specifically proposed for OCT to OCTA translation. 
Adjacent GAN~\cite{li2020deep} incorporates three adjacent OCT B-scan images and outputs the middle OCTA B-scan image. Furthermore, we include two methods designed for paired image-to-image translation. VQ-I2I \cite{chen2022eccv} leverages vector quantization techniques within a GAN framework, achieving state-of-the-art performance in paired image-to-image translation tasks. Palette \cite{saharia2022palette} builds upon conditional diffusion models, with the target domain images as conditions and noised source domain images as inputs. Among the six compared methods, pix2pix 3D is trained on 3D volumes, while the remaining methods are trained on 2D B-scan images. 

From the results, we observe that pix2pix 3D generally outperformed other methods, indicating the significance of inter-slice connections in providing more global information compared to a single B-scan image. The latest approaches, VQ-I2I and Palette, achieve higher performances than previous methods. VQ-I2I benefits from its powerful backbone architectures, although it comes with a high computational cost when handling 3D input data. 
Palette utilizes diffusion models but had longer inference times. In contrast to these methods, our proposed TransPro method can handle 3D inputs, which leverages enhanced global contextual information. Meanwhile, it maintains negligible inference times compared to diffusion models. Moreover, TransPro incorporates vessel and contextual guidance from VPG and HCG modules, leading to significant improvements in the OCTA-3M dataset compared to other methods in both B-scan images and projection maps. Notably, in Table~\ref{table1}, our method demonstrates relative enhancements over the pix2pix 3D model by 40.8\% in VDE and 14.0\% in VDC metrics, respectively.

\subsubsection{Experimental results on OCTA-6M}
We further conduct the experiments of our proposed TransPro and six comparative methods on the OCTA-6M dataset, as shown in Table~\ref{table2}. It is observed that the quality of translated images in this dataset is generally lower compared to the OCTA-3M dataset. This discrepancy may be attributed to the lower resolution of images due to the larger field of view in the OCTA-6M dataset. 

Regarding the results among the compared methods, the outcomes align with those obtained on the OCTA-3M dataset, where the performance is positively correlated with the utilization of global information by the models. Our proposed TransPro method achieves the best overall performance among all the compared methods, except for a slightly lower PSNR value compared to the Pix2pix 3D model. Notably, the remarkable improvements in the VDE and VDC metrics demonstrate the effectiveness of our method in enhancing the quality of vascular structures.
These results highlight the strengths of TransPro in leveraging global contextual information and effectively enhancing the quality of vascular structures in OCTA images, even in challenging datasets such as OCTA-6M. Despite the lower resolution and potentially more complex images in this dataset, our method consistently outperforms the other compared methods, showcasing its effectiveness in improving the translation quality of OCT to OCTA images.

\subsubsection{Visualization results}
To evaluate the results qualitatively, we provide some visualization examples of \mia{B-scan images} and projection maps translated by different methods. \mia{For the B-scan images, we select two examples from OCTA-3M and OCTA-6M datasets and compare the translated results of Pix2pix 3D, Adjacent GAN, and our TransPro method in Figure~\ref{fig3.0}. All methods are capable of learning the general outline and the texture of OCTA B-scans, albeit with variations in some details.  The results obtained from Pix2pix 3D appear blurrier and exhibit discontinuities in certain areas, while the Adjacent GAN tends to predict the B-scan images with more white tissues, which may obscure the essential vessels. To further demonstrate the performance qualitatively, we compute the MAE for each image denoted in red boxes. Overall, our proposed TransPro method achieves the best translation results of OCTA B-scans in both qualitative and quantitative evaluation.}
For the projection maps, we choose four examples from OCTA-3M and OCTA-6M datasets respectively and compare the results of our proposed TransPro with Pix2pix 3D and Adjacent GAN methods. From the zoomed-out figures in Figure~\ref{fig3}, we find that the Pix2pix 3D model generates some artifacts to mimic the texture of vessels while the outputs of the Adjacent GAN model suffer from white stripes due to the limited intra-slice information of the 2D model. Compared to the ambiguous images translated from other methods, our method produces high-quality OCTA projection images with vivid vessels and background tissues.

\subsection{Ablation Study}

\begin{figure*}[t]
\includegraphics[width=.9\textwidth]{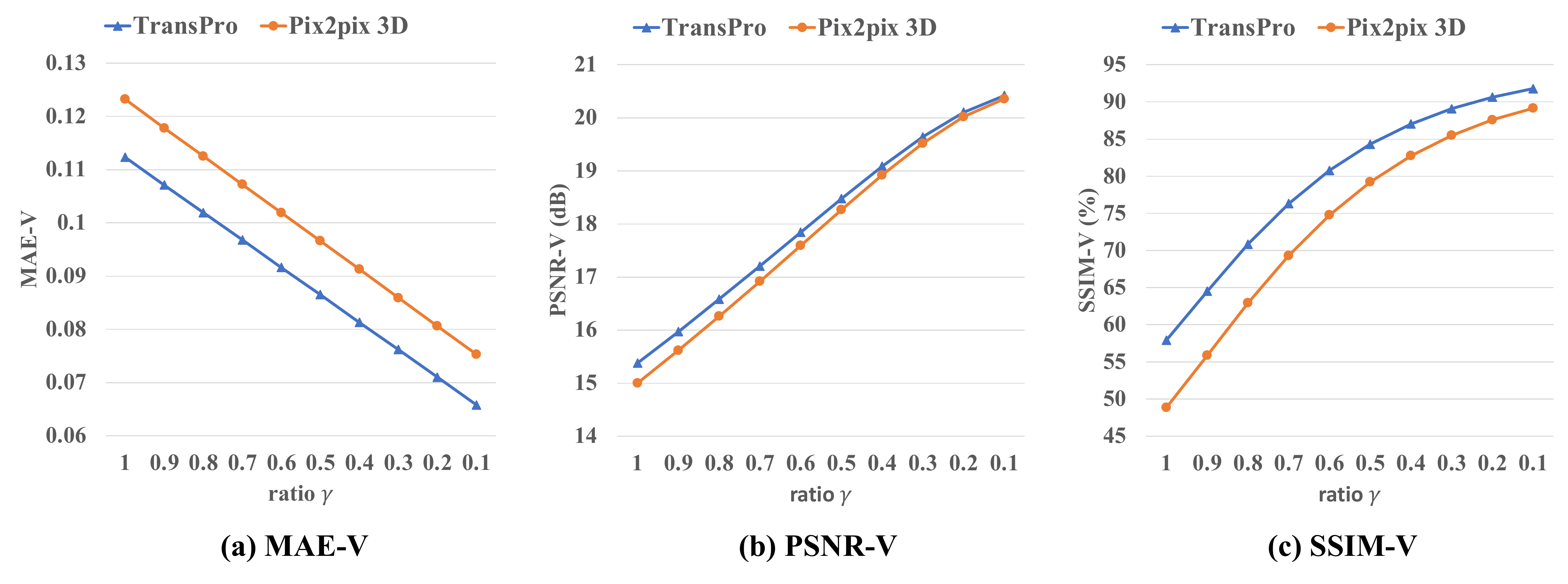}
\centering
\vspace{-3mm}
\caption{Evaluation of translated vascular quality on OCTA projection in OCTA-3M~\cite{li2020ipn}. The non-vascular pixels are multiplied by a ratio $\gamma$ from 1.0 to 0.1. The three evaluation metrics, \ie, (a) MAE-V, (b) PSNR-V and (c) SSIM-V, are computed on processed OCTA projection maps. The x-axis refers to the coefficient multiplied to non-vascular pixels, and the y-axis refers to the results of corresponding metrics.}
\label{ratio}
\end{figure*}
\begin{figure*}[t]
\includegraphics[width=.9\textwidth]{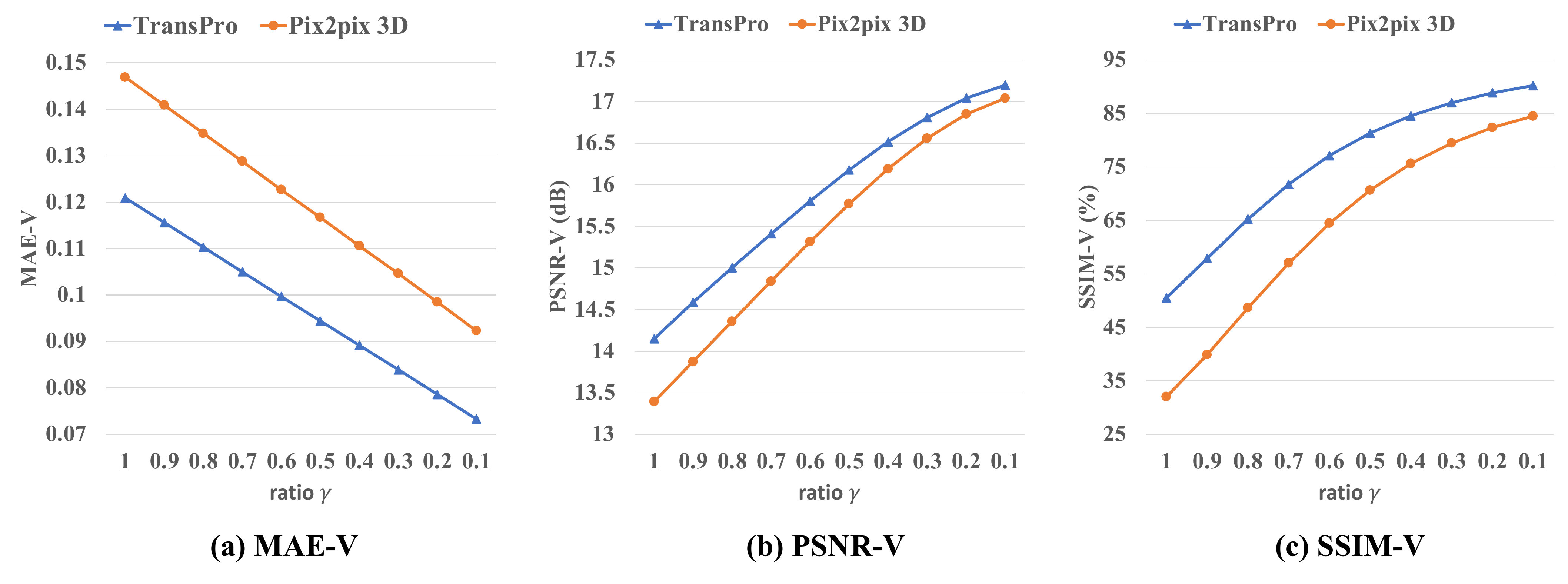}
\centering
\vspace{-3mm}
\caption{Evaluation of translated vascular quality on OCTA projection in OCTA-6M~\cite{li2020ipn}. The non-vascular pixels are multiplied by a ratio $\gamma$ from 1.0 to 0.1. The three evaluation metrics, \ie, (a) MAE-V, (b) PSNR-V and (c) SSIM-V, are computed on processed OCTA projection maps. The x-axis refers to the coefficient multiplied to non-vascular pixels, and the y-axis refers to the results of corresponding metrics.}
\vspace{-4mm}
\label{ratio_2}
\end{figure*}

\subsubsection{Effectiveness of VPG and HCG} 
\begin{figure}[t]
\includegraphics[width=.48\textwidth]{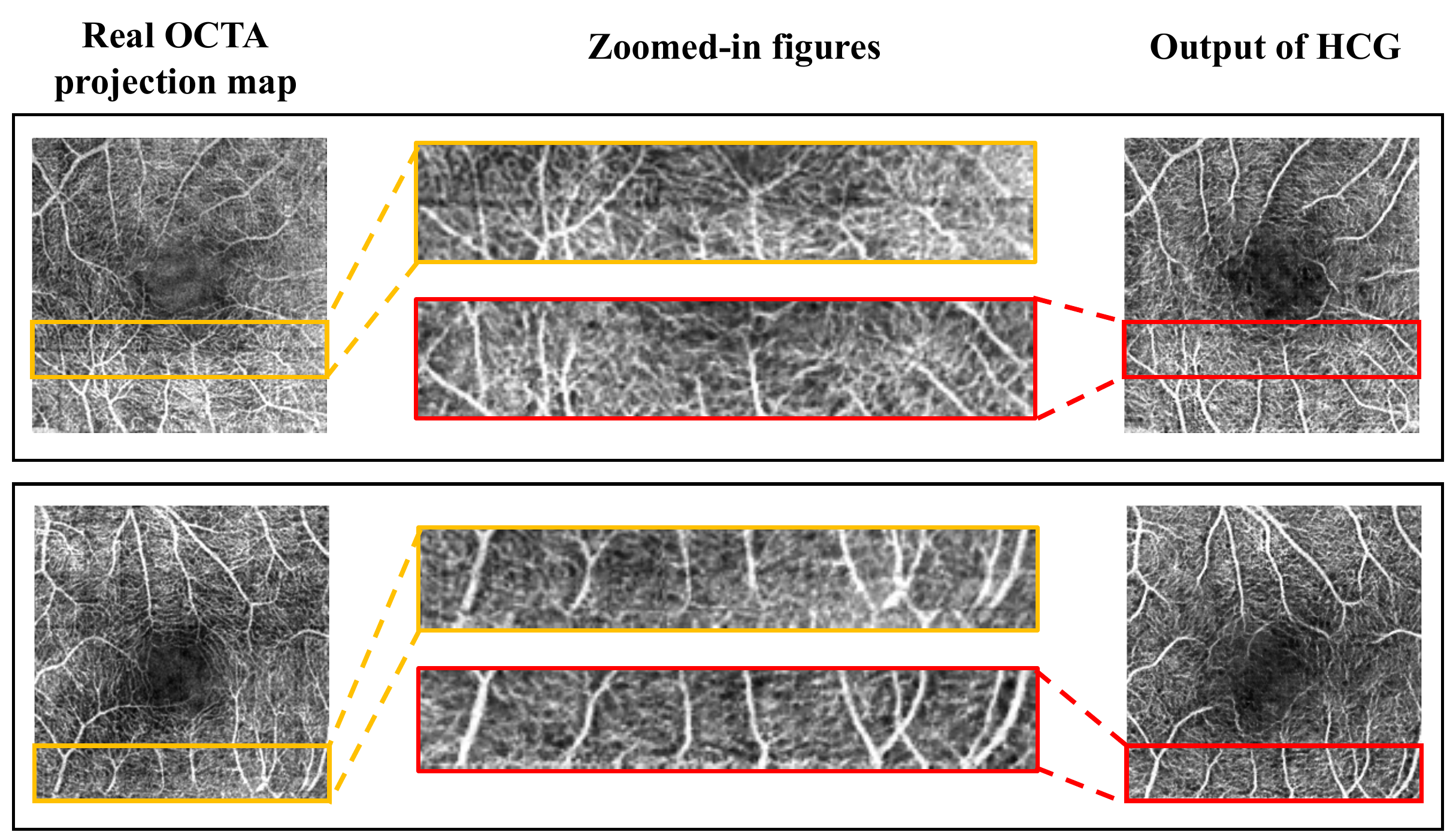}
\centering
\vspace{-2mm}
\caption{\mia{Motivation of our proposed Heuristic Contextual Guidance (HCG) module. We compare the real OCTA projection maps and the outputs of the HCG model. Two examples show the vessel discontinuity in the real OCTA projection maps while the HCG output images recover the connection.}}
\vspace{-4mm}
\label{fig_hcg}
\end{figure}

To validate the respective effectiveness of VPG and HCG components, we conduct additional experiments by separately adding these two modules to the 3D GAN baseline model. Table \ref{table3} displays the results on OCTA-3M and OCTA-6M datasets. In general, we can find that VPG and HCG modules bring improvement to the quality of B-scan images as the metrics of MAE, PSNR and SSIM evaluating OCTA B-scan images become better after adding corresponding components. For the quality of projection maps of translated OCTA images, the vessel density error (VDE) and vessel density correlation (VDC) metrics indicate the effectiveness of VPG and HCG modules on vascular areas. In particular, the VPG module shows significant improvement on VDC metric because it provides the semantic vascular segmentation map which increases the similarity of overall vascular structures. \mia{The HCG module acts as a regularizer, effectively addressing the overfitting issue caused by specific patterns found in real OCTA projection maps, such as the vessel discontinuity illustrated in Figure~\ref{fig_hcg}. By employing the convolution operations on 2D OCTA projection maps in the HCG module, each output pixel incorporates information from nearby pixels. This integration of contextual information significantly enhances the model's overall performance and leads to a notable reduction in the VDE metric, particularly when applied to the OCTA-3M dataset.}
Moreover, when the VPG and HCG modules are applied simultaneously on the baseline model, \ie, our proposed TransPro framework, the performance obtains prominent gains compared to the results with either VPG or HCG module. Therefore, we conclude that VPG and HCG modules offer their own effectiveness on the OCT to OCTA image translation task and their effects are additive when the two modules are active in TransPro method.

\begin{table*}
\centering

\caption{Ablation studies on different losses of VPG and HCG modules on OCTA-3M dataset~\cite{li2020ipn}. The MAE, PSNR (dB), and SSIM (\%) are computed on OCTA volumes, while the MAE-V, PSNR-V (dB), SSIM-V (\%), VDE, and VDC (\%) are computed on OCTA projection maps. ``$\downarrow$'' refers to the lower the better and ``$\uparrow$'' refers to the higher the better.}
\resizebox{\linewidth}{!}{
\begin{tabular}{c|c|ccc|ccccc}
\toprule
Module & Loss type & MAE~$\downarrow$ & PSNR(dB)~$\uparrow$ & SSIM(\%)~$\uparrow$ & MAE-V$\downarrow$ & PSNR-V(dB)$\uparrow$ & SSIM-V(\%)$\uparrow$ & VDE$\downarrow$ & VDC(\%)$\uparrow$ \\

\hline
\multirow{3}{*}{VPG} & MSE & 0.0819 & 31.78 & 86.99 & 0.0734 & 20.29 & 89.55 & \mia{0.1610} & \mia{69.29} \\
 & BCE & 0.0794 & \textbf{33.38} & 87.68 & 0.0669 & 20.35 & 91.25 & \mia{0.1673} & \mia{68.84} \\
 & L1 & \textbf{0.0782} & 32.56 & \textbf{88.22} & \textbf{0.0658} & \textbf{20.42} & \textbf{91.97} & \textbf{\mia{0.1304}} & \textbf{\mia{74.41}}\\

\hline

\multirow{3}{*}{HCG} & MSE & 0.0798 & 32.14 & 87.53 & 0.0702 & 20.10 & 90.49 & \mia{0.1452} & \mia{72.12}\\
 & SSIM & 0.0846 & 31.62 & 87.18 & 0.0797 & 19.31 & 87.61 & \mia{0.1352} & \mia{73.78}\\
 & L1 & \textbf{0.0782} & \textbf{32.56} & \textbf{88.22} & \textbf{0.0658} & \textbf{20.42} & \textbf{91.97} & \textbf{\mia{0.1304}} & \textbf{\mia{74.41}}\\

\bottomrule
\end{tabular}}
\label{table4}
\end{table*}
\begin{table*}
\centering

\caption{Ablation studies on task-balanced hyperparameters $\alpha$ \mia{and $\beta$} on OCTA-3M dataset~\cite{li2020ipn}. The MAE, PSNR (dB), and SSIM (\%) are computed on OCTA volumes, while the MAE-V, PSNR-V (dB), SSIM-V (\%), VDE, and VDC (\%) are computed on OCTA projection maps. ``$\downarrow$'' refers to the lower the better and ``$\uparrow$'' refers to the higher the better.}
\resizebox{\linewidth}{!}{
\begin{tabular}{c|ccc|ccccc}
\toprule
Hyparameters & MAE~$\downarrow$ & PSNR(dB)~$\uparrow$ & SSIM(\%)~$\uparrow$ & MAE-V$\downarrow$ & PSNR-V(dB)$\uparrow$ & SSIM-V(\%)$\uparrow$ & VDE$\downarrow$ & VDC(\%)$\uparrow$ \\

\hline
$\alpha$, \mia{$\beta$}=1 & 0.0815 & 32.32 & 88.22 & 0.0716 & 20.30 & 90.30 & \mia{0.1450} & \mia{71.71} \\
$\alpha$, \mia{$\beta$}=3 & 0.0839 & 31.43 & 87.84 & 0.0762 & 18.05 & 88.85 & {\mia{0.1360}} & \mia{73.27} \\
$\alpha$, \mia{$\beta$}=5 & \textbf{0.0782} & \textbf{32.56} & \textbf{88.22} & \textbf{0.0658} & \textbf{20.42} & \textbf{91.97} & \textbf{\mia{0.1304}} & \textbf{\mia{74.41}}\\
$\alpha$, \mia{$\beta$}=7 & 0.0813 & 32.27 & 87.40 & 0.0692 & 20.20 & 91.10 & \mia{0.1398} & \mia{72.38}\\
$\alpha$, \mia{$\beta$}=9 & 0.0831 & 31.59 & 87.08 & 0.0727 & 19.75 & 89.83 & \mia{0.1418} & \mia{72.06} \\

\bottomrule
\end{tabular}}
\label{table5}
\end{table*}
\subsubsection{Evaluation of translated vascular quality}
\label{Ablation-2}
To evaluate the quality of translated vessels in OCTA projection maps, we further calculate vessel-weighted MAE, PSNR, and SSIM metrics by adjusting the ratio of pixel weights between vessel pixels and non-vessel pixels according to the vessel segmentation mask. We summarize the results in line charts for different ratios in OCTA-3M and OCTA-6M datasets, as shown in Figure \ref{ratio} and \ref{ratio_2}, respectively. 
Specifically, we first obtain the pixel-wise vascular labels of OCTA projection maps in the test set which are annotated by ophthalmologists. Then, for the non-vessel pixels in the segmentation mask, we multiply the OCTA projection maps by a ratio $\gamma \in \{1.0, 0.9, ..., 0.1\}$ to gradually decrease the weights of non-vessel pixels from 1.0 to 0.1. The weights of vessel pixels remain at 1.0. We apply the weights adjusting for all translated OCTA projection maps and the ground-truth images and compute MAE-V, PSNR-V, and SSIM-V metrics between two sets of projection maps. In consequence, as the weights of vascular pixels increase, MAE-V drops linearly and the metric of PSNR-V and SSIM-V gradually improve for both OCTA-3M and OCTA-6M datasets. 
In particular, the PSNR-V and SSIM-V results of TransPro method can achieve over 20 dB and 90\% for the OCTA-3M dataset and over 17 dB and 90\% for the OCTA-6M dataset, respectively, when the weights of non-vessel pixels reduce to 0.1. In addition, by comparing our method to Pix2pix 3D method on both two datasets, the results indicate that our method can achieve higher quality in both non-vessel regions (when the ratio is high) and vessel regions (when the ratio is low) than pix2pix 3D model.

\subsubsection{Types of losses in VPG and HCG modules}
\label{Ablation-3}

To validate the suitability of L1 loss for the VPG (Equation~\ref{vesselseg_loss}) and HCG (Equation~\ref{context_loss}) modules, we conduct experiments to compare it with alternative loss functions. The translation results of L1 loss outperform other types of loss, as shown in Table~\ref{table4}.

For the VPG module, we employ Mean Squared Error (MSE) loss and Binary Cross Entropy (BCE) loss to enforce consistency between the segmentation predictions of the translated OCTA projection map and the ground-truth projection map. Since the output of the vessel segmentation model is normalized between 0 and 1, the MSE loss yields smaller values than the L1 loss, resulting in a weaker constraint and poorer translation results. The BCE loss demonstrates comparable performance with L1 loss, achieving a higher PSNR value for OCTA volumes while slightly reducing other metrics.

For the HPG module, we utilize MSE loss and Structure Similarity (SSIM) loss to minimize differences between two OCTA projection maps. The loss function serves as an image reconstruction metric by measuring the dissimilarities between two images. From the results, we find that MSE loss exhibits slightly inferior performance to L1 loss because it leads to blurred translated images. SSIM loss is defined to measure image similarity in terms of luminance, contrast, and structure. However, due to the noise and ambiguity inherent in OCTA projection maps, SSIM loss faces challenges in accurately reflecting image similarity.

\subsubsection{Hypaparameters}
\label{Ablation-4}

We complete a series of additional experiments to evaluate the impact of the task-balance hyperparameters $\alpha$ \mia{and $\beta$} in Equation~\ref{total_loss} for OCTA translation. \mia{The VPG and HCG losses exhibit comparable magnitudes during training. To maintain computational feasibility of the hyperparameter search, we assign equal weights to these losses.} We evaluate the performance of our method on the OCTA-3M dataset and analyze the results, which are presented in Table~\ref{table5}. The results demonstrate that our proposed method consistently enhances the overall performance across a wide range of $\alpha$ \mia{and $\beta$} values. We identify that the best performance is achieved when $\alpha$ \mia{and $\beta$} are set to 5, indicating the importance of selecting an appropriate value to balance the main task and two auxiliary tasks.

\begin{figure}[t]
\includegraphics[width=.49\textwidth]{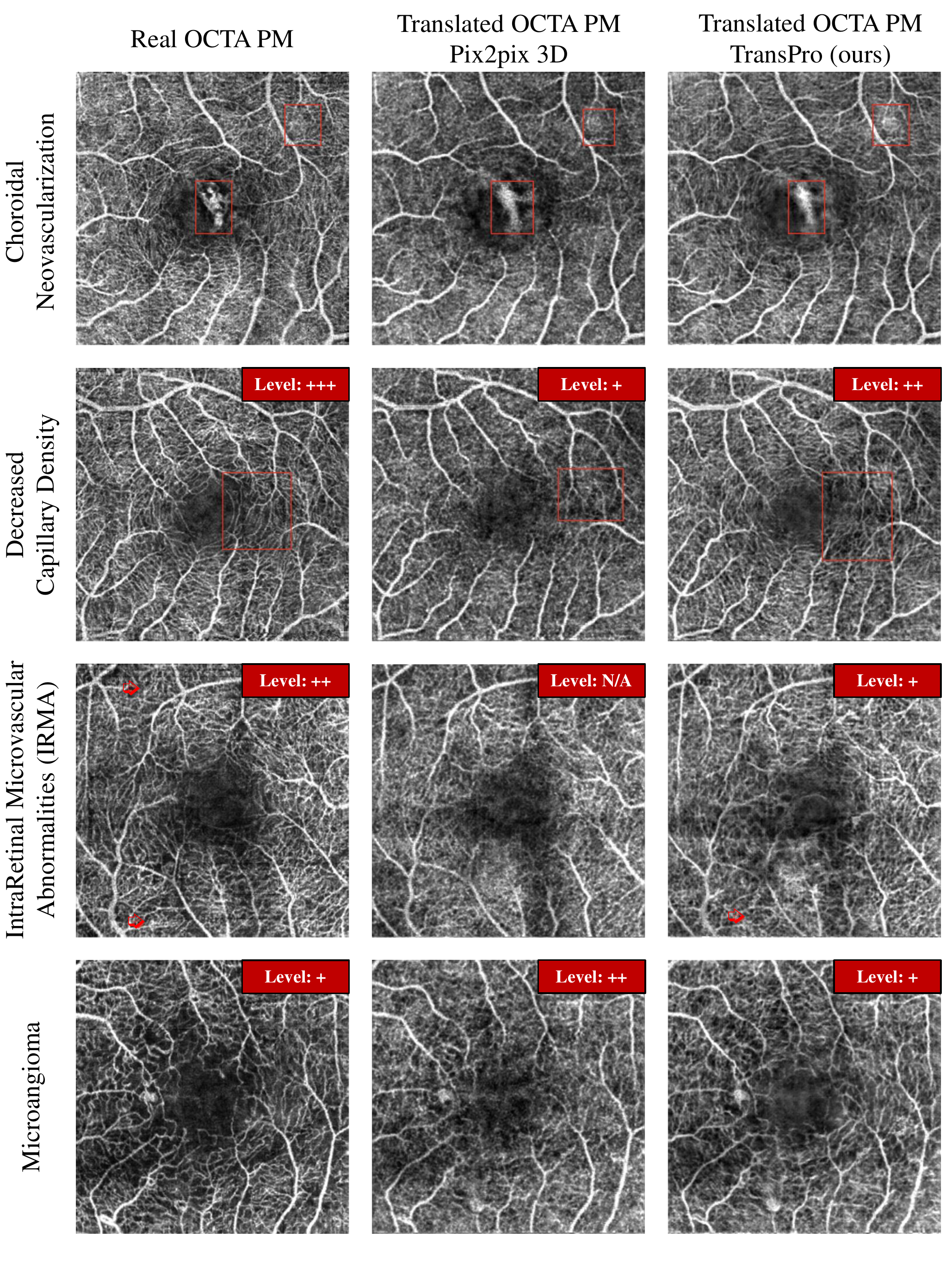}
\centering
\vspace{-2mm}
\caption{\mia{The diseased patterns of translated OCTA projection maps annotated by an experienced ophthalmologist. Four different patterns are displayed, including choroidal neovascularization, decreased capillary density, intraretinal microvascular abnormalities (IRMA), and microangioma. Each pattern is presented alongside the corresponding real OCTA image, the translation results from the Pix2pix 3D method, and our TransPro method. Levels of disease patterns are annotated from “+” to “+++”, where the number of “+” denotes the degree of severity.}}
\vspace{-4mm}
\label{fig_patterns}
\end{figure}
\mia{
\subsubsection{Abnormal pattern identification and disease recognition}
\label{Ablation-5}
In addition to the quantitative and qualitative analysis of the translated OCTA images, we cooperate with ophthalmologists to analyze the abnormal patterns and diseases in the translated OCTA images, such that the potential practical applications of deep learning-based OCT to OCTA image translation method are evaluated. We conduct two tasks, which are abnormal pattern identification and disease recognition.

For the abnormal pattern identification task, we collaborate with an experienced ophthalmologist who examined both real and translated OCTA projection maps. We summarize four types of abnormal patterns including choroidal neovascularization, decreased capillary density, intraretinal microvascular abnormalities (IRMA), and microangioma, as shown in Figure~\ref{fig_patterns}. For each type of pattern, we compare the real OCTA images, the translation results from Pix2pix 3D method, and our TransPro method. Our findings reveal that major diseased patterns are preserved in the translated OCTA projection maps, albeit with some ambiguity in the details of fine vessel areas. Moreover, compared to the Pix2pix 3D method, our TransPro generates OCTA projection maps that are more similar to the real ones in terms of diseased levels and locations, as annotated in Figure~\ref{fig_patterns}.

For the disease recognition task, we conduct an evaluation where an experienced ophthalmologist classifies diabetic retinopathy (DR) and choroidal neovascularization (CNV) images among 50 samples from both real and translated OCTA projection maps. The disease recognition results of DR and CNV on the real and translated images are identical, achieving a 100\% accuracy rate. Although this experiment encompasses a limited number of disease types and samples, the high consistency and accuracy of the OCTA projection maps translated by our proposed TransPro method highlight their quality and potential for practical applications in OCTA translation.}

However, although the generated OCTA projection maps provide valuable insights for disease diagnosis, they do not faithfully reconstruct the finer vascular details. Therefore, our proposed TransPro method cannot completely replace the necessity for precise OCTA scanning. It is potential that future developments in deep-learning-based OCT to OCTA image translation algorithms will yield more accurate and higher fidelity results.

\section{Conclusion}
\label{5:con}
In this paper, a novel framework for 3D OCT to OCTA image translation, termed TransPro, is proposed. TransPro is mainly driven by two key insights: leveraging two different input views and improving vessel area quality in the translated OCTA images. The first insight is inspired by a crucial observation that the OCTA projection map is generated by averaging pixel values from its corresponding B-scans along the Z-axis. As a result, we propose a hybrid generative architecture with a novel Heuristic Contextual Guidance (HCG) module, that ensures the consistency of the translated OCTA images in both B-scan and projection views. To achieve the second insight, we propose a novel Vessel Promoted Guidance (VPG) module, which enhances the attention on retinal vessels.
Experimental results on public datasets demonstrate that TransPro achieves outstanding performance. TransPro is a versatile 3D image generation model that can be applied to a wide range of computer vision tasks, such as data augmentation, semantic segmentation, and domain adaptation. Future work will explore the use of TransPro in additional image generation applications.

\section{Acknowledgements}
This work is supported in part by 
Foshan HKUST Projects (Grants FSUST21-HKUST10E and FSUST21-HKUST11E), in part by a grant from
Science and Technology Planning Project of Guangdong Province (No. 2023A0505030004) and projects of 
Hetao Shenzhen-Hong Kong Science and Technology Innovation Cooperation Zone (HZQB-KCZYB2020083).

\bibliographystyle{model2-names.bst}\biboptions{authoryear}
\bibliography{refs}

\end{document}